\nonstopmode

\newcommand{\re}{{\mathrm{Re}\,}}
\newcommand{\im}{{\mathrm{Im}\,}}
\newcommand{\eV}{\U{eV}}
\newcommand{\mul}{\cdot}
\newcommand{\bohr}{\, a_0}
\newcommand{\Hartree}{\U{E_{\mathit h}}}
\newcommand{\angstrom}{\U{\hbox{\AA}}}
\newcommand{\degree}{{}^{\circ}}
\newcommand{\invcm}{\U{cm^{-1}}}
\newcommand{\ie}{i.e.{}}
\newcommand{\eg}{e.g.{}}
\newcommand{\etal}{\textit{et al.}}
\newcommand{\U}[1]{\,{\rm{#1}}}
\newcommand{\I}[1]{_{\mathrm{#1}}}
\newcommand{\imag}{{\rm i}}
\newcommand{\euler}{\mathrm e}
\newcommand{\Sum}{\sum\limits}
\newcommand{\Int}{\int\limits}
\newcommand{\mod}{\,\mathrm{mod}\,}
\newcommand{\transpose}{{}^{\textrm{\scriptsize T}}}
\newcommand{\differential}{\>\mathrm d}
\newcommand{\bra}[1]{\left<\right.\!#1\!\left.\right|}
\newcommand{\ket}[1]{\left|\right.\!#1\!\left.\right>}
\newcommand{\CFtBr}{CF$_3$Br}
\newcommand{\expectval}[1]{\left<#1\right>}
\newcommand{\cleb}[2]{C(#1 ; #2)}
\newcommand{\E}[1]{\times 10^{#1}}
\newcommand{\dfrac}[2]{{\displaystyle{#1}\over\displaystyle{#2}}}

\documentclass[prb,twocolumn,showpacs,nopreprintnumbers,superscriptaddress]{revtex4}
\usepackage{bm,bbm,amssymb,subeqnarray,graphicx}
\usepackage[nativepdf,bookmarks,bookmarksopen,bookmarksnumbered,raiselinks,%
pdftitle={Rotational molecular dynamics of laser-manipulated bromotrifluoromethane
studied by x-ray absorption},%
pdfauthor={Christian Buth, Robin Santra},%
pdfsubject={Chemical physics},%
pdfkeywords={ab initio, laser alignment, x rays, photoabsorption, cross section, %
polarization dependence, perturbation theory, nonperturbative, %
laser-matter interaction, CF3Br, bromotrifluoromethane}]{hyperref}

\begin{document}
\title{Rotational molecular dynamics of laser-manipulated bromotrifluoromethane
studied by x-ray absorption}
\thanks{The following article appeared in J.{} Chem.{} Phys.{} \textbf{129},
134312 (2008) (12~pages) and may
be found at \href{http://dx.doi.org/10.1063/1.2987365}
                        {dx.doi.org/10.1063/1.2987365}.
Copyright~2008 American Institute of Physics.
This article may be downloaded for personal use only.
Any other use requires prior permission of the author and the American
Institute of Physics.}
\author{Christian Buth}
\thanks{Present address: Department of Physics and Astronomy, Louisiana State
University, Baton Rouge, Louisiana~70803, USA}
\email{christian.buth@web.de}
\affiliation{Argonne National Laboratory, Argonne, Illinois~60439, USA}
\author{Robin Santra}
\affiliation{Argonne National Laboratory, Argonne, Illinois~60439, USA}
\affiliation{Department of Physics, University of Chicago, Chicago,
Illinois 60637, USA}
\date{October 14, 2008}

\begin{abstract}
We present a computational study of the rotational molecular dynamics of
bromotrifluoromethane~(\CFtBr{}) molecules in gas phase.
The rotation is manipulated with an offresonant, $800 \U{nm}$~laser.
The molecules are treated as rigid rotors.
Frequently, we use a computationally efficient linear rotor model
for~\CFtBr{} which we compare with selected results for full
symmetric-rotor computations.
The expectation value~$\expectval{\cos^2 \vartheta}(t)$ is discussed.
Especially, the transition from impulsive to adiabatic alignment,
the temperature dependence of the maximally achievable alignment and
its intensity dependence are investigated.
In a next step, we examine
resonant x-ray absorption as an accurate tool to study laser manipulation
of molecular rotation.
Specifically, we investigate the impact of the x-ray pulse duration on the
signal (particularly its temporal resolution), and study the temperature
dependence of the achievable absorption.
Most importantly, we demonstrated that using picosecond x-ray pulses,
one can accurately measure the expectation value~$\expectval{\cos^2
\vartheta}(t)$ for impulsively aligned \CFtBr{}~molecules.
We point out that a control of the rotational dynamics opens up a novel
way to imprint shapes onto long x-ray pulses on a picosecond time scale.
For our computations, we determine the dynamic polarizability tensor
of~\CFtBr{} using \emph{ab initio} molecular linear-response
theory in conjunction with wave function models of increasing sophistication:
coupled-cluster singles~(CCS), second-order approximate coupled-cluster
singles and doubles~(CC2), and coupled-cluster singles and
doubles~(CCSD).
\end{abstract}

%
%
%

\preprint{arXiv:0809.0146}
\maketitle

\section{Introduction}

We consider an ensemble of bromotrifluoromethane~(\CFtBr{}) molecules
subjected to the light of an optical, nonresonant laser.
Due to a dynamic second order Stark effect, the electric field of the
laser exerts a net torque on the molecular axis with the largest
polarizability.
It is directed in such a way that it aligns the molecular axis with the
linear polarization axis of the laser and thus aligns the molecules.
The laser intensity is too low to excite or ionize the
molecules.~\cite{Stapelfeldt:AM-03,Seideman:NA-05,Santra:SF-07}
In combination with x~rays, we have a so-called two-color problem;
the laser can be used in a pump-probe-way preceding the x~rays
or simultaneously with them.
The former case leads to impulsive (transient) alignment where a short pulse
creates a rotational wavepacket which evolves freely afterwards and periodically
goes through brief periods of alignment;
the latter case typically is used in conjunction with long laser pulses
leading to adiabatic alignment where the molecular rotation
follows the laser pulse envelope.
Laser alignment has been extensively studied experimentally [see,
\eg, Ref.~\onlinecite{Stapelfeldt:AM-03}].
It was explained in the adiabatic case by Friedrich and
Herschbach~\cite{Friedrich:AT-95} and in the case of pulsed lasers
inducing rotational wavepacket dynamics leading to transient alignment
by Seideman.~\cite{Seideman:RE-95}
The laser alignment of symmetric rotors was studied by
Hamilton~\etal{}~\cite{Hamilton:AS-05}
Recently, we formulated a theory for laser-aligned symmetric-top molecules
which are probed \emph{in situ} by x~rays;
it was applied to study bromine molecules.~\cite{Buth:LA-08}
In a joint experimental and theoretical investigation, the alignment
of~\CFtBr{} was studied for various
conditions.~\cite{Santra:SF-07,Peterson:XR-08}

In this work, we would like to deepen the theoretical understanding of the
x-ray probe of molecules.
Especially, we are interested in trends.
Resonant x-ray absorption is sensitive to molecular alignment, if the
resonance has certain symmetry properties.
Then selection rules can be established which lead to
a different absorption for perpendicular and parallel x-ray polarization
vectors with respect to the polarization axis of the laser.
However, the absorption of x~rays above the edge is not sensitive
to molecular alignment.
Rotational dynamics is conventionally specified in terms of
the expectation value~$\expectval{\cos^2 \vartheta}(t)$.
This value is a theoretical quantity.
Experimentally, one typically has resorted to the Coulomb explosion
method to study alignment.~\cite{Stapelfeldt:AM-03,Seideman:NA-05}
This method relies on the intricate nature of molecular strong-field
ionization.~\cite{Ellert:MA-99,Pavicic:DM-07}
The x-ray absorption technique uses the simpler one-photon interaction and
thus is physically simpler.
We address the following questions in this paper.
In what way is~$\expectval{\cos^2 \vartheta}(t)$ accessible from x-ray
absorption measurements in a cross-correlation experiment?
What is the achievable temporal resolution?
We see our work in conjunction with the fascinating possibilities offered by
upcoming ultrafast x-ray sources.~\cite{LCLS:CDR-02,Tanaka:SC-05,Borland:SA-05,%
Altarelli:TDR-06}
They will provide intriguing direct insights into the
molecular rotational dynamics in the time domain.
Additionally, our research lays the foundation for molecular imaging
of laser-aligned molecules using x-ray diffraction.~\cite{Ryu:DS-03,Buth:US-up}

In addition to the analytical value of the x-ray absorption probe,
laser-induced molecular alignment opens up a way to control x-ray absorption
in an ultrafast (picosecond) way.
This means that by controlling the alignment of the molecules in a gas sample,
we are able to control how much flux of an incident x-ray pulse is transmitted
by the gas.
By exploiting the available advanced optical technology to shape laser pulses,
this offers a novel route to imprint shapes on the long x-ray pulses
($\sim 100 \U{ps}$) from third-generation light sources.

This paper is structured as follows.
Section~\ref{sec:theory} discusses the x-ray absorption cross section
of \CFtBr{}~molecules for a two-level electronic structure model
and its relation to~$\expectval{\cos^2 \vartheta}(t)$;
a reduction of the cross section to experimentally easily accessible
quantities;
and a model for a temperature dependent rotational period.
Computational details are given in Sec.~\ref{sec:compdet}.
There, the dynamic dipole polarizability of~\CFtBr{} and its anisotropy
are also computed.
Results are presented in Sec.~\ref{sec:results} for molecular rotational
dynamics represented by~$\expectval{\cos^2 \vartheta}(t)$ and the x-ray absorption
technique as well as the control of x~rays by molecular rotation.
Conclusions are drawn in Sec.~\ref{sec:conclusion}.
In the Appendix, we show that symmetry-breaking effects are small
in~\CFtBr{} that modify the selection rules for resonant x-ray absorption.

Our equations are formulated in atomic units,~\cite{Szabo:MQC-89}
where $1 \U{hartree} = 1 \Hartree$ is the unit of energy,
$1 \, t_0$~is the unit of time,
and $1 \U{bohr} = 1 \bohr$ is the unit of length.
The Boltzmann constant~$k\I{B}$ is unity and
$1 \Hartree = 3.15775 \E{5} \U{K}$~is the unit of temperature.
Intensities are given in units of~$1 \Hartree \> t_0^{-1} \, a_0^{-2}
= 6.43641 \times 10^{15} \U{W \, cm^{-2}}$ and
electric polarizabilities are measured in~$1 \, e^2 \, a_0^2 \, \Hartree^{-1}
= 1.64877725 \E{-41} \U{C^2 \, m^2 \, J^{-1}}$.

\section{Theory}
\label{sec:theory}
\subsection{Electronic structure model for \protect\CFtBr{}}
\label{sec:elstructmod}

\begin{figure}
  \includegraphics[clip,width=5cm]{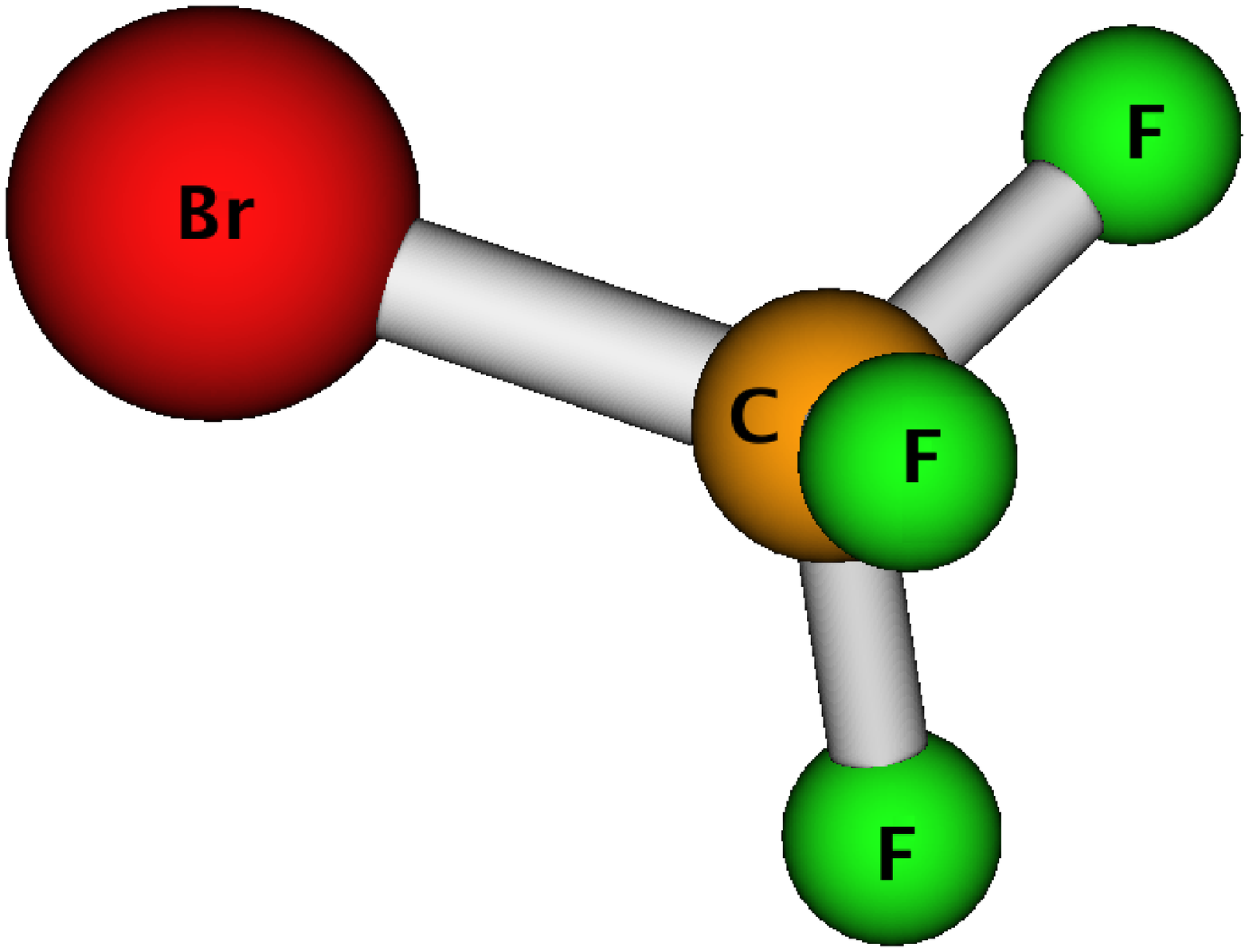}
  \caption{(Color online) The molecular structure
           of~\CFtBr{}.~\cite{footnote2}\protect\nocite{Schaftenaar:MO-00,%
           dalton:pgm-05,Wilson:GBS-99,basislib:02-02-06}}
  \label{fig:geomCF3Br}
\end{figure}

Our investigations are based on our theory of x-ray absorption
by laser-aligned symmetric-top molecules in Ref.~\onlinecite{Buth:LA-08}.
There, we derive the following formula for the instantaneous x-ray absorption
cross section at time~$t$ of a molecule manipulated by a laser
field:~\cite{footnote1}
\begin{equation}
  \label{eq:xsect}
  \begin{array}{rcl}
    \displaystyle \sigma(t) &=& \displaystyle 4 \pi \, \alpha \, \omega\I{X}
      \  \im \biggl[ \Sum_{J,J',K,M} \varrho_{JJ'}^{(KM)}(t)
      \Sum_{J'',K'',M'', i''} \\
    &&\displaystyle{} \hspace{-1em} \times
      \frac{(\vec d^{\>\prime\,*}_{0i''}\mul\vec s_{J'KM, J''K''M''}^{\>\prime})
            (\vec d^{\>\prime\,*}_{i''0}\mul\vec s_{J''K''M'',JKM}^{\>\prime})}
      {E_{i''} - E_0 + E_{J''K''} - E_{JK} - \omega\I{X} - \imag \, \Gamma / 2}
      \biggr] \; .
  \end{array}
\end{equation}
We denote the fine-structure constant by~$\alpha$ and the x-ray photon energy
by~$\omega\I{X}$.
The density matrix of the laser-only problem in terms of symmetric-rotor states
is~$\varrho_{J J'}^{(KM)}(t)$.
We use~$J$, $K$, $M$ (also with accents) to denote the quantum numbers of
symmetric-top wave functions~\cite{Zare:AM-88} and $i$ (also with accents) to
numerate electronic states.~\cite{Buth:LA-08}
The electronic ground state is indicated by~$i=0$.
We assume the Born-Oppenheimer approximation and consider nuclear and electronic
degrees of freedom to be totally separated from each
other.~\cite{Szabo:MQC-89,Buth:LA-08}
The matrix elements of the complex conjugate of the dipole operator~\cite{footnote1} in the
body-fixed reference frame are indicated by~$\vec d^{\>\prime}_{0i''}$.
With~$\vec s_{J'KM, J''K''M''}^{\>\prime}$, we denote the matrix elements of the x-ray
polarization vector in the body-fixed frame in terms of symmetric-top wave functions.
The primes on the vectors indicate that they are given in terms of the spherical
basis instead of the Cartesian basis.~\cite{Rose:ET-57,Zare:AM-88}
Energies of symmetric-rotor states and electronic states are represented
by~$E_{JK}$, $E_{J''K''}$ and $E_0$, $E_{i''}$, respectively.
The finite decay width of core-excited states due to Auger decay and x-ray
fluorescence is described by~$\Gamma$.~\cite{Als-Nielsen:EM-01,Thompson:XR-01}

\begin{figure}
  \begin{center}
    a)~~\includegraphics[clip,width=3.5cm]{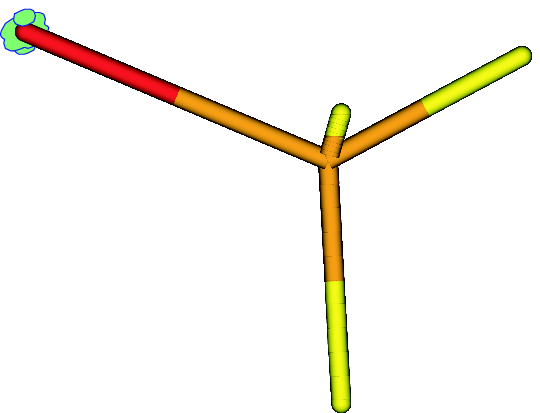}
    \qquad b)~\includegraphics[clip,width=3.5cm]{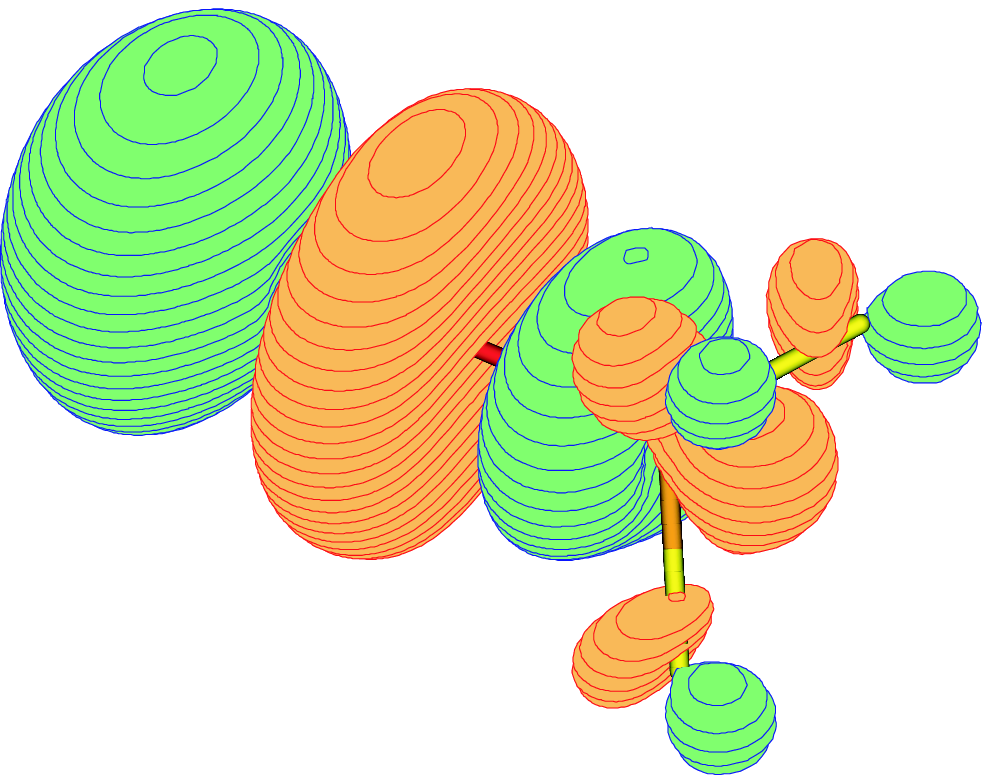}
    \caption{(Color online) The two orbitals included in the two-level model
             for the electronic structure of a \CFtBr{}~molecule from which
             the x-ray absorption cross section is
             determined.~\cite{footnote2}\protect\nocite{Schaftenaar:MO-00,%
             dalton:pgm-05,Wilson:GBS-99,basislib:02-02-06}
             a)~The Br$\,1s$~orbital;
             b)~the $\sigma^*$~antibonding molecular orbital.}
    \label{fig:twoorbs}
  \end{center}
\end{figure}

The molecule \CFtBr{} (the molecular structure in Fig.~\ref{fig:geomCF3Br})
has an isolated resonance below the bromine $K$~edge in the x-ray
absorption spectrum [confer Fig.~1 of Ref.~\onlinecite{Peterson:XR-08}].
Assuming that the x-ray photon energy is tuned to this resonance, a
simplification of the electronic structure of the molecule to two levels is
reasonable.
The pre-edge resonance is assigned to the transition from the Br$\,1s$~orbital
to the $\sigma^*$~antibonding molecular orbital.
See Fig.~\ref{fig:twoorbs} for a graphical representation
of the two orbitals used.

Resonant photoabsorption is sensitive to the alignment of \CFtBr{}~molecules, because
the orbitals are glued to the molecular frame and thus rotate with it.
Bromotrifluoromethane has C$_{3v}$~symmetry;
the Br$\,1s$ and $\sigma^*$~orbitals have $A_1$~symmetry.
The transition is mediated by the Cartesian dipole matrix
element~$\bra{\mathrm{Br}\,1s} \hat{\vec d} \ket{\sigma^*}$ with~$\hat{\vec d}
= (a, b, c)\transpose$ in the molecule-fixed reference frame.
Due to the vanishing integrals rule,~\cite{Bishop:GT-73,Atkins:MQM-04,Buth:LA-08}
the direct product of representations~$\Gamma^{* \, \mathrm{Br}\,1s}
\otimes \Gamma^i\otimes \Gamma^{\sigma^*}$ with $i \in \{a, b, c\}$
needs to contain the totally symmetric representation such that
$\bra{\mathrm{Br}\,1s} (\hat{\vec d})_i \ket{\sigma^*} \neq 0$, \ie, the
transition~$\mathrm{Br}\,1s \to \sigma^*$ is allowed via~$(\hat{\vec d})_i$.
Using the product table of the C$_{3v}$~point group,~\cite{Atkins:MQM-04}
we find that the $c$~component of~$\hat{\vec d}$ is totally symmetric but
the components~$a$ and $b$ are not.
This gives rise to a dichroism with respect to the figure axis (C--Br)
of~\CFtBr{}: absorption for parallel x-ray polarization vector, no
absorption for perpendicular x~ray polarization vector.
In the Appendix, we show that this result holds in spite of potential
symmetry-breaking effects in~\CFtBr{}.

For our two-level model,
we have~$\bra{\mathrm{Br}\,1s} \hat{\vec d} \ket{\sigma^*}
= d\I{c} \, \vec e_c = d\I{c} \, \vec e_0 \equiv \vec d^{\>\prime}_{10}$
with the unit vectors~$\vec e_c$ and $\vec e_0$ along the $c$ and the
$m=0$~axes, respectively.~\cite{Kroto:MR-75,Rose:ET-57,%
Zare:AM-88,Buth:LA-08}
We neglect the dependence of the denominator in Eq.~(\ref{eq:xsect}) on
rotational quantum numbers.
This is justified because, in view of the large inner-shell decay widths,
rotational states remain unresolved in x-ray absorption.
Let us decompose the fraction into a real part and an imaginary part
as follows:
\begin{equation}
  \label{eq:2levfrac}
  F\I{R} + \imag \, F\I{I} \equiv \frac{1}{E_1 - E_0 - \omega\I{X}
    - \imag \, \Gamma / 2} \; .
\end{equation}
We tune the x~rays to the resonance energy, \ie, $\omega\I{X} = E_{1} - E_0$.
Thus, we find from Eq.~(\ref{eq:2levfrac}) $F\I{R} = 0$ and $F\I{I} =
\frac{2}{\Gamma}$.
The resonant cross section~(\ref{eq:xsect}) of a two-level system becomes:
\begin{equation}
  \label{eq:xsecttwo}
  \begin{array}{rcl}
    \displaystyle \sigma_2(t) &=& \displaystyle
      8\pi \, \alpha \, \omega\I{X} \, \frac{|d_c|^2}{\Gamma}
      \Sum_{J,J',K,M} \  \re \varrho_{J J'}^{(KM)}(t) \\
    &&\displaystyle{} \times \Sum_{J'',K'',M''} s_{c, J'KM,
      J''K''M''}^{\>\prime} \> s_{c, J''K''M'',JKM}^{\>\prime} \; ,
  \end{array}
\end{equation}
with~$s_{c,J'KM, J''K''M''}^{\>\prime} \equiv (\vec s_{J'KM, J''K''M''}^{\>\prime})_c$
and  $s_{c, J''K''M'',JKM }^{\>\prime} \equiv (\vec s_{J''K''M'',JKM  }^{\>\prime})_c$.

The initial, thermal distribution over rotational states is sensitive
to the nuclear spin statistics.~\cite{Kroto:MR-75}
The three fluorine atoms in~\CFtBr{}, are~$^{19}_{\phantom{0}9}$F, the
only stable fluorine isotope, with nuclear spin
quantum number~$I = \frac{1}{2}$.~\cite{Mills:QU-88}
In this case, the statistical weights of the symmetric rotor
are given by~\cite{Kroto:MR-75}
\begin{equation}
  g_I(J,K) = g_{\frac{1}{2}}(K) = \cases{ 4 & ; $K \mod 3   =  0$ \cr
                                          2 & ; $K \mod 3 \neq 0 \; .$}
\end{equation}
In most of the computations of this paper, we use a linear model for~\CFtBr{}.
In this case, the CF$_3$~portion of~\CFtBr{} is assumed to be a pseudo-atom
which is distinct from a bromine atom.
Thus the nuclear statistical weight is in this case:
\begin{equation}
  g_I(J) = 1 \; .
\end{equation}
There is no quantum number~$K$ for a linear rotor.~\cite{Kroto:MR-75}

\subsection{Experimentally accessible quantities}

The instantaneous cross section~(\ref{eq:xsect}) is not easily experimentally
accessible.
However, we can use it to devise quantities which can be measured more readily.
The laser pulse is imprinted on the cross section of the two-level
model~$\sigma_{2,i}(t)$ [Eq.~(\ref{eq:xsecttwo})].
We determine it for parallel ($i = \parallel$) and perpendicular ($i = \perp$)
x-ray and laser polarization vectors.
Additionally, we consider the cross section without
laser~$\sigma_{2,\mathrm{th}}(t)$, \ie, a thermal ensemble.
The x~rays are characterized by the photon flux~$J\I{X}(t)$.
The cross correlation of the cross section and the x-ray pulse is defined by
\begin{equation}
  \label{eq:crosscorr}
  P_i(\tau) = \Int_{-\infty}^{\infty} \sigma_{2,i}(t) \, J\I{X}(t-\tau)
    \differential t \; , \qquad i = \parallel, \perp, \mathrm{th} \; .
\end{equation}
It represents the total probability of x-ray absorption
for a time delay of~$\tau$ between the laser and the x-ray pulse.
We define following ratios of cross correlations:
\begin{subeqnarray}
  \label{eq:crossratios}
  \slabel{eq:crossparperp}
  R\I{\parallel / \perp      }(\tau) &=& \frac{P_{\parallel}(\tau)}
    {P_{\perp      }(\tau)} \; , \\
  R\I{\parallel / \mathrm{th}}(\tau) &=& \frac{P_{\parallel}(\tau)}
    {P_{\mathrm{th}}(\tau)} \; , \\
  R\I{\perp     / \mathrm{th}}(\tau) &=& \frac{P_{\perp    }(\tau)}
    {P_{\mathrm{th}}(\tau)} \; .
\end{subeqnarray}
The ratios were studied experimentally in
Ref.~\onlinecite{Santra:SF-07,Peterson:XR-08}.

Let us assume ultrashort x-ray pulses which can be modeled by a
$\delta$~distribution as follows:
\begin{equation}
  \label{eq:deltaflux}
  J\I{X}(t - \tau) = J\I{X,0} \, \delta(t - \tau) \; .
\end{equation}
Inserting this flux into Eq.~(\ref{eq:crosscorr}) yields~$P_i(\tau) =
J\I{X,0} \, \sigma_{2,i}(\tau)$.
When the ratios~(\ref{eq:crossratios}) are formed, the constant~$J\I{X,0}$
cancels and the ratios are given by the ratios of the participating cross
sections.
Hence, in the limit of ultrashort x-ray pulse duration,
Eq.~(\ref{eq:crossratios}), goes over into the new ratios
\begin{subeqnarray}
  \label{eq:crossratioscross}
  \slabel{eq:crossparperpcross}
  r\I{\parallel / \perp      }(\tau) &=& \frac{\sigma_{2,\parallel}(\tau)}
    {\sigma_{2,\perp      }(\tau)} \; , \\
  \slabel{eq:crossparthcross}
  r\I{\parallel / \mathrm{th}}(\tau) &=& \frac{\sigma_{2,\parallel}(\tau)}
    {\sigma_{2,\mathrm{th}}(\tau)} \; , \\
  \slabel{eq:crossperpthcross}
  r\I{\perp     / \mathrm{th}}(\tau) &=& \frac{\sigma_{2,\perp    }(\tau)}
    {\sigma_{2,\mathrm{th}}(\tau)} \; .
\end{subeqnarray}

\subsection{Relation between the cross section and~$\expectval{\cos^2
\vartheta}(t)$}
\label{sec:relcrosscos}

We set out from the coupling matrix elements in
Eq.~(\ref{eq:xsect}) reading~\cite{Buth:LA-08,footnote1}
\begin{equation}
  \label{eq:cplmat}
  \vec d^{\>\prime\,*}_{0i''} \mul \vec s^{\>\prime}_{J'KM,J''K''M''}
    = \bra{J'KM, 0} \hat{\vec d} \mul \vec e\I{X} \ket{J''K''M'', i''} \; .
\end{equation}
The x-ray polarization vector is~$\vec e\I{X}$ and
$\hat{\vec d}$~denotes the dipole operator.
Both vectors are Cartesian vectors in the laboratory reference frame.
The scalar product on the left hand side of Eq.~(\ref{eq:cplmat}) is
formed in the spherical basis.~\cite{Rose:ET-57,Zare:AM-88}
With~$\ket{J''K''M'', i''}$ we denote the direct product of the
symmetric-rotor state~$\ket{J''K''M''}$ and the electronic
state~$\ket{i''}$.

Let us assume the two-level model of Sec.~\ref{sec:elstructmod};
the only dipole vector~$\vec d_{01}$ and the polarization
vector~$\vec e\I{X}$ are real vectors.
The scalar product on the right hand side of Eq.~(\ref{eq:cplmat}) becomes
\begin{equation}
  \begin{array}{rl}
    &\displaystyle \bra{J'KM} \vec d_{01}^* \mul \vec e\I{X} \ket{J''K''M''} \\
    \displaystyle =& \displaystyle |\vec d_{01}| \, \bra{J'KM} \cos \vartheta
      \ket{J''K''M''} \; .
  \end{array}
\end{equation}
We use~$|\vec e\I{X}| = 1$ and $\vartheta$ is the angle between the two
vectors~$\vec d_{01}$ and $\vec e\I{X}$.
We choose $\vec e\I{X} = \vec e_z$ along laser polarization axis.
The dipole vector~$\vec d_{01}$---here given in the laboratory frame---points
along the C--Br~molecular axis;
in the body-fixed reference frame the C--Br~axis was chosen to be the $c$~axis
[Sec.~\ref{sec:elstructmod}].
Thus, $\vartheta$ is the Euler angle between the $z$ and $c$-axes of
the space-fixed and molecule-fixed frame, respectively.
The dependence of the denominator in Eq.~(\ref{eq:xsect}) on
rotational quantum numbers is disregarded as in Eq.~(\ref{eq:2levfrac}).
The sum over intermediate symmetric-rotor states can then be eliminated
totally yielding~$\bra{J'KM} \cos^2 \vartheta \ket{JKM}$ in the numerator of
Eq.~(\ref{eq:xsect}).
The remaining summation forms a trace with these matrix elements and the density
matrix:
\begin{equation}
  \expectval{\cos^2 \vartheta}(t) = \Sum_{J,J',K,M} \varrho_{JJ'}^{(KM)}(t)
    \bra{J'KM} \cos^2 \vartheta \ket{JKM} \; .
\end{equation}
Finally, we arrive at the cross section of the two-level
model~(\ref{eq:xsecttwo}) which clearly exhibits the relation
to~$\expectval{\cos^2 \vartheta}(t)$ as follows:
\begin{equation}
  \label{eq:xsecttwocospar}
  \sigma_{2,\parallel}(t) = 8\pi \, \alpha \, \omega\I{X} \, \frac{|d_c|^2}{\Gamma}
    \expectval{\cos^2 \vartheta}(t) \; .
\end{equation}

Assuming an x-ray polarization vector perpendicular to the laser-polarization
axis yields
\begin{equation}
  \label{eq:xsecttwocosperp}
  \sigma_{2,\perp}(t) = 4\pi \, \alpha \, \omega\I{X} \, \frac{|d_c|^2}{\Gamma}
    [1 - \expectval{\cos^2 \vartheta}(t)] \; .
\end{equation}
The prefactor of~$1 - \expectval{\cos^2 \vartheta}(t)$ in~$\sigma_{2,\perp}(t)$
is only half the prefactor of~$\expectval{\cos^2 \vartheta}(t)$
in~$\sigma_{2,\parallel}(t)$ because there are two perpendicular directions
to the laser-polarization axis compared to only one for parallel
polarization vectors.
The cross section of a thermal ensemble can be inferred easily from
Eq.~(\ref{eq:xsecttwocospar}) by letting~$\expectval{\cos^2 \vartheta}(t)
= \frac{1}{3}$ which leads to
\begin{equation}
  \label{eq:xsecttwocostherm}
  \sigma_{2,\mathrm{th}}(t) = \frac{8\pi}{3} \, \alpha \, \omega\I{X} \,
    \frac{|d_c|^2}{\Gamma} \; .
\end{equation}
Given the formulas for the cross sections~(\ref{eq:xsecttwocospar}),
(\ref{eq:xsecttwocosperp}), and (\ref{eq:xsecttwocostherm}), we can
write explicit expressions for the ratios in Eq.~(\ref{eq:crossratioscross})
as follows:
\begin{subeqnarray}
  \label{eq:crossratiostwo}
  \slabel{eq:crossparperptwo}
  r\I{\parallel / \perp      }(\tau) &=& \frac{2 \expectval{\cos^2 \vartheta}(\tau)}
    {1 - \expectval{\cos^2 \vartheta}(\tau)} \; , \\
  \slabel{eq:crossparthtwo}
  r\I{\parallel / \mathrm{th}}(\tau) &=& 3\expectval{\cos^2 \vartheta}(\tau) \; , \\
  \slabel{eq:crossperpthtwo}
  r\I{\perp     / \mathrm{th}}(\tau) &=& \frac{3}{2} \, [1 -
    \expectval{\cos^2 \vartheta}(\tau)] \; .
\end{subeqnarray}

\subsection{Thermal average of the rotational period}
\label{sec:thaverp}

The character of molecular alignment---impulsive to adiabatic see,
\eg, Fig.~\ref{fig:cos2t_7} and the ensuing discussion---crucially
depends on the rotational temperature~$T$.~\cite{Stapelfeldt:AM-03,%
Seideman:NA-05}
Conventionally, the rotational period of a linear rotor is defined by~$T\I{RP}
= \frac{1}{2B}$ with the rotational constant~$B$.
The regimes of molecular alignment from impulsive to adiabatic are
distinguished by comparing~$T\I{RP}$
with the duration of the aligning laser pulse~$\tau\I{L}$.
Yet, as $T\I{RP}$ is independent of~$T$, the classification thus does not
vary with~$T$, despite the changed alignment characteristics.

\newcommand{\crulefill}{\hrulefill}

\begin{table*}
  \centering
  \begin{ruledtabular}
    \begin{tabular}{rlcccccc}
                &     & \multicolumn{2}{c}{\crulefill CCS\crulefill}
                      & \multicolumn{2}{c}{\crulefill CC2\crulefill}
                      & \multicolumn{2}{c}{\crulefill CCSD\crulefill} \\
    \hfill  Basis set\hfill\hfill & $\omega\I{L}$ [$\Hartree$]
      & $\bar\alpha$ [a.u.] & $\Delta\alpha$ [a.u.]
      & $\bar\alpha$ [a.u.] & $\Delta\alpha$ [a.u.]
      & $\bar\alpha$ [a.u.] & $\Delta\alpha$ [a.u.] \\
      \hline
          cc-pVDZ & 0     & 25.54 & 14.81 & 26.11 & 15.97 & 25.05 & 14.80\\
                  & 0.057 & 25.71 & 15.09 & 26.32 & 16.32 & 25.24 & 15.12\\[3pt]
      aug-cc-pVDZ & 0     & 36.60 & 11.67 & 38.83 & 13.57 & 37.20 & 12.16\\
                  & 0.057 & 36.93 & 11.89 & 39.27 & 13.90 & 37.60 & 12.43\\[3pt]
      aug-cc-pVTZ & 0     & 37.71 & 10.91 & 40.05 & 12.78 &  &           \\
                  & 0.057 & 38.07 & 11.10 & 40.52 & 13.08 &  &
    \end{tabular}
  \end{ruledtabular}
  \caption{Dynamic average dipole polarizability~$\bar\alpha(\omega\I{L})$
           and dynamic dipole polarizability anisotropy~$\Delta
           \alpha(\omega\I{L})$ of a \CFtBr{}~molecule in the field of a
           laser with photon energy~$\omega\I{L}$ determined with
           coupled-cluster linear response methods for several basis
           sets.~\cite{Buth:LA-08}
           Goebel and Hohm~\cite{Goebel:DD-97} obtained the experimental
           values~$\bar\alpha(0 \Hartree) = 41.23 \U{a.u.}$ and
           $\Delta \alpha(0 \Hartree) = 11.00 \U{a.u.}$}
  \label{tab:polarizability}
\end{table*}

We derive a thermally averaged rotational period for a linear molecule
which does not exhibit this shortcoming.
Therefore, we equate the energy expression of the
\emph{classical} linear rotor of an angular frequency~$\omega_J$ with the
\emph{quantum mechanical} rotational energy corresponding to the angular
momentum~$J$ as follows:
\begin{equation}
  E_J = \frac{1}{2} \, I\I{B} \,  \omega_J^2 = B J (J+1) \; .
\end{equation}
The moment of inertia for the rotation around axis~$B$ is~$I\I{B} =
\dfrac{1}{2B}$.~\cite{Kroto:MR-75}
This leads to the rotational period
\begin{equation}
  T_J = \frac{\pi}{\sqrt{B \, E_J}} \; .
\end{equation}
A thermal ensemble of rotational states of temperature~$T = \frac{1}{\beta}$
consequently has the average rotational period:
\begin{equation}
  \label{eq:T_rot}
  T\I{th} = \frac{1}{Z} \Sum_{J=1}^{\infty} T_J \> g_I(J) \> (2J+1)
    \, \euler^{-\beta \, E_J} \; .
\end{equation}
The nuclear statistical weight is given by~$g_I(J)$
[Sec.~\ref{sec:elstructmod}].
The partition function reads
\begin{equation}
  \label{eq:partition}
  Z = \Sum_{J=1}^{\infty} g_I(J) \, (2J+1) \, \euler^{-\beta \, E_J} \; .
\end{equation}
The term for the rotational ground state~$J=0$ is explicitly excluded in the
sums of Eqs.~(\ref{eq:T_rot}) and (\ref{eq:partition}) because a molecule
in the $J=0$~state does not rotate and thus~$T_0 = \infty$.

\section{Computational details}
\label{sec:compdet}

Bromotrifluoromethane~(\CFtBr) [Fig.~\ref{fig:geomCF3Br}] is a prolate
symmetric-top molecule of C$_{3v}$~symmetry.
We use the structural data from Taylor~\cite{Taylor:RS-54} where the
bond lengths are specified as~$\rm |C\,F| = 1.328 \angstrom$ and
$\rm |C\,Br| = 1.918 \angstrom$ and the bond angles are specified
as~$\angle\rm (F\,C\,F) = 108.42 \degree$ and $\angle\rm (F\,C\,Br)
= 110.52 \degree$.
The rotational constants~$A = 0.191205 \invcm$ and $B = 0.069834 \invcm$ were
obtained with the \emph{ab initio} quantum chemistry program package
\textsc{dalton}.~\cite{dalton:pgm-05,footnote3}%
\nocite{Mills:QU-88,Rosman:IC-98}
In the linear rotor approximation, we set~$A = 0$ which implies no rotations
around the figure axis.~\cite{Kroto:MR-75}
For \CFtBr{}, the rotational times are~$T\I{A} = 1 / (2A) = 87.2 \U{ps}$ and
$T\I{B} = 1 / (2B) = 238.8 \U{ps}$.
Here, $B$ is the crucial constant because it corresponds to
a rotation about an axis perpendicular to the figure axis which is the
only one influenced by the laser.~\cite{footnote4}
The other constant~$A$ creates only a substructure of the energy
levels of the thermal density matrix of the symmetric-top
molecule.~\cite{Buth:LA-08}

We compute the average dynamical dipole polarizability~$\bar\alpha(\omega\I{L})$
of the molecule and its anisotropy~$\Delta \alpha(\omega\I{L})$
in the field of the aligning laser.~\cite{Maroulis:CA-04}
The polarizability anisotropy is needed to describe the impact of the laser on
the alignment later on;
the average polarizability merely causes a constant energy shift of the
Hamiltonian which can be neglected for our purposes.~\cite{Buth:LA-08}
Both quantities are determined in fixed-nuclei approximation using
\textsc{dalton}.~\cite{dalton:pgm-05}
We harness molecular linear-response theory in conjunction with
wave function models of increasing sophistication:
coupled-cluster singles~(CCS), second-order approximate coupled-cluster
singles and doubles~(CC2), and coupled-cluster singles and
doubles~(CCSD).~\cite{Christiansen:ID-98}
The cc-pVDZ [correlation consistent polarized valence double zeta] and
aug-cc-pV$X$Z [augmented cc-pV$X$Z] (for $X={}$D, T [D:~double,
T:~triple]) basis sets~\cite{Wilson:GBS-99,basislib:02-02-06} are used.

We examine the static case, $\omega\I{L} = 0 \Hartree$, and the dynamic case,
$\omega\I{L} = 0.057 \Hartree$, for the photon energy of an $800 \U{nm}$
Ti:sapphire laser.
Our data are compiled in Table~\ref{tab:polarizability}.
A moderately rapid convergence with respect to the basis set quality is
observed for the individual methods.
We were not able to determine data for the combination of~CCSD
with aug-cc-pVTZ due to insufficient computational resources.
Our best results for~$\omega\I{L} = 0 \Hartree$ are determined using the
CC2 method with the aug-cc-pVTZ basis set.
We note a satisfactory agreement of our values to those of Goebel and
Hohm,~\cite{Goebel:DD-97} who obtain an average static
polarizability of~$41.23 \U{a.u.}$ and an anisotropy of~$11.00 \U{a.u.}$.
In our computations, we use the value of~$\Delta \alpha(0.057 \Hartree)
= 12.00 \U{a.u.}$ for the polarizability anisotropy.

The computations in this paper were carried out with the program
\textsc{alignmol} of the \textsc{fella} package.~\cite{fella:pgm-V1.3.0}
The initial density matrix of a thermal ensemble of rotational states is
represented using symmetric-rotor functions up to an angular momentum
denoted by~$J\I{max}$ which was chosen high enough for the respective contribution
to the partition function of the thermal ensemble to be less
than~\cite{Buth:LA-08}~$10^{-6}$.
For temperatures lower than~$6 \U{K}$, $N\I{eq} = 25$ equations of motions are
used for each angular momentum to describe the time-evolution of the rotational
density matrix of the molecules interacting with the laser;
for higher temperatures, $N\I{eq} = 50$ is used.
Unless otherwise stated, we assume Gaussian laser and x-ray pulses;
their durations (FWHM) are denoted by~$\tau\I{L}$ and $\tau\I{X}$, respectively.
The wave packet propagation takes place over the time interval
which is subdivided into~$N\I{t}$ time steps.
The number was always chosen sufficiently high with respect to~$\tau\I{L}$ and
$\tau\I{X}$ for the numerical integration of the differential equations to be
converged.
The number of x-ray pulses which are used for cross correlation
calculations~$N\I{X}$ is sufficiently high to ensure smooth
curves in the respective graphs of this paper.

\section{Results and discussion}
\label{sec:results}
\subsection{Laser alignment}
\label{sec:lasalign}

Let us examine~$\expectval{\cos^2 \vartheta}(t)$ which is typically used
to quantify molecular alignment.~\cite{Stapelfeldt:AM-03,Seideman:NA-05}
Here, $\vartheta$ represents the polar Euler angle which describes the
tilt of the $c$~axis---identical to the figure axis in our case---of
the molecule-fixed coordinate system with respect to the laser polarization
axis, which is the $z$~axis of the space-fixed coordinate
system.~\cite{Kroto:MR-75,Zare:AM-88}
The quantity~$\expectval{\cos^2 \vartheta}(t)$ can be inferred, \eg, with
the help of the Coulomb explosion
technique.~\cite{Stapelfeldt:AM-03,Seideman:NA-05}
To this end, the molecule is highly ionized with an intense laser pulse.
It breaks up into fragments that subsequently undergo Coulomb explosion.
The location where the fragments hit the detector plate is used to
infer~$\expectval{\cos^2 \vartheta}(t)$.

\begin{figure}
  \includegraphics[clip,width=\hsize]{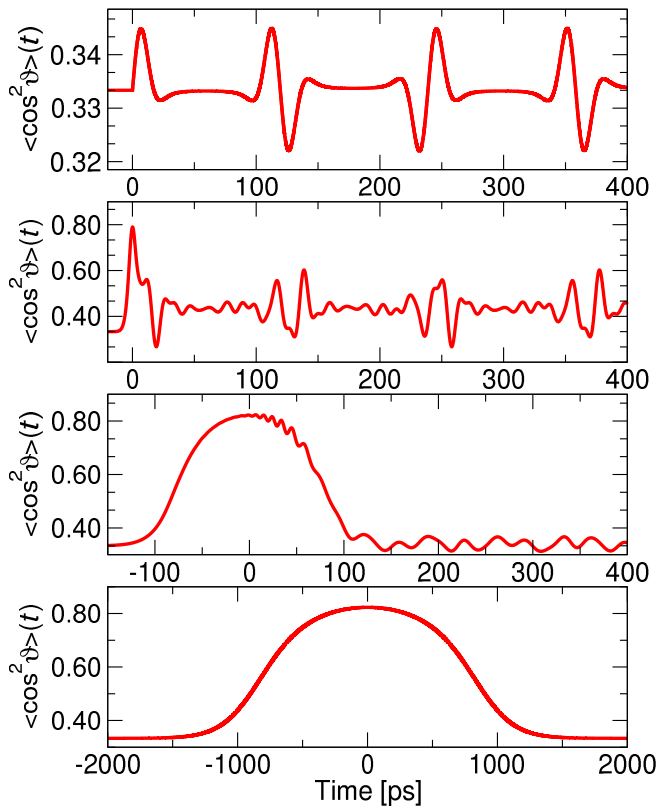}
  \caption{(Color online) The time evolution of~$\expectval{\cos^2
           \vartheta}(t)$ due to molecular rotation is shown for four
           temporal regimes of the laser pulse:
           a)~impulsive alignment, $\tau\I{L} = 50 \U{fs}$;
           b)~short-pulse intermediate regime (still almost impulsive),
           $\tau\I{L} = 10 \U{ps}$;
           c)~long-pulse intermediate regime (quasi-adiabatic),
           $\tau\I{L} = 95 \U{ps}$;
           d)~adiabatic alignment, $\tau\I{L} = 1 \U{ns}$.
           Other computational parameters are~$T=1 \U{K}$,
           and $I\I{L,0} = 10^{12} \U{\frac{W}{cm^2}}$.}
  \label{fig:cos2t_7}
\end{figure}

In Fig.~\ref{fig:cos2t_7}, we investigate molecular dynamics for four regimes
of laser-pulse durations.
These regimes are characterized by the ratio of the laser-pulse duration
and the thermally averaged rotational constant [Sec.~\ref{sec:thaverp}].

First, there is pure transient alignment [top panel of Fig.~\ref{fig:cos2t_7}];
the molecules rotate freely after an initial ``kick'' with a laser pulse
which is short compared to the rotational response time~$T\I{th}$
[Eq.~(\ref{eq:T_rot})] of the molecule.
The short laser pulse induces a brief period of alignment.
After well-defined periods of time, half revivals occur around~$120 \U{ps}$
and around~$360 \U{ps}$.
At such occurrences, there is a brief period of alignment followed
by a brief period of antialignment.
The molecule undergoes full revivals around~$240 \U{ps}$ exhibiting the reverse
sequence---antialignment followed by alignment---compared with half
revivals.~\cite{Hamilton:AS-05}
Antialignment means that the molecules are aligned perpendicular to the
laser polarization axis.
Then, the expectation value~$\expectval{\cos^2 \vartheta}(t)$ is suppressed with
respect to the value of a thermal ensemble.

Second, the short-pulse intermediate regime is displayed in the upper middle
panel of Fig.~\ref{fig:cos2t_7}.
Here, the molecular dynamics is still almost impulsive.
The purely impulsive character of the top panel of the figure, however,
is distorted due to a longer interaction with the laser pulse during which
rotational motion takes place.

Third, the long-pulse intermediate regime is shown in the lower middle
panel of Fig.~\ref{fig:cos2t_7};
the molecular response is almost adiabatic.
Only slight nonadiabaticities occur, \eg, the ripples on top of the
peak.
Moreover, the molecular ensemble does not fully return to a thermal
distribution after the laser pulse has ended.~\cite{footnote5}
This causes beating of excited states which leads to the oscillations
beyond~$100 \U{ps}$.

Fourth, in the lowest panel of Fig.~\ref{fig:cos2t_7}, we display
pure adiabatic alignment.
The rotational response follows closely the laser pulse shape.
After the pulse, the ensemble has returned to the initial thermal distribution.
We observe that the peak alignment of the molecules is much larger
in the adiabatic case compared with the impulsive case in the top panel.
This occurs due to the fact that a larger number of rotational states can
be excited in the wave packet during a longer laser
pulse which, in turn, facilitates a stronger localization of the molecule, \ie,
alignment.~\cite{Stapelfeldt:AM-03,Seideman:NA-05}

\begin{figure}
  \includegraphics[clip,width=\hsize]{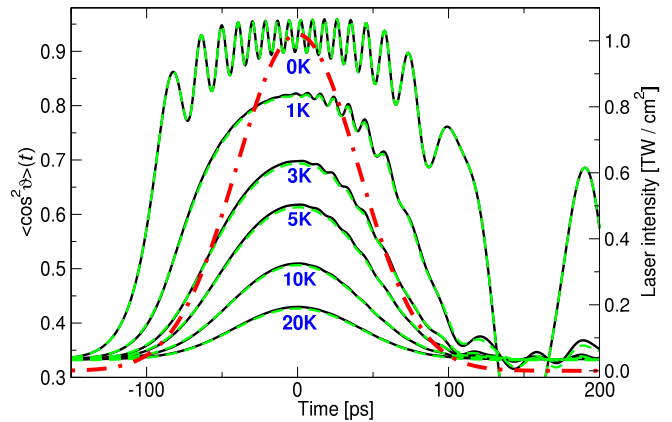}
  \caption{(Color) Influence of the temperature on the rotational
           dynamics of~\CFtBr{}.
           The time evolution of~$\expectval{\cos^2 \vartheta}(t)$ is
           depicted for a linear rotor by the solid (black) curves for a range of
           temperatures given in blue.
           For comparison, we show results from a symmetric-rotor computation as
           dashed (green) curves.
           The Gaussian laser pulse with~$\tau\I{L} = 95 \U{ps}$ and
           $I\I{L,0} = 10^{12} \U{\frac{W}{cm^2}}$ is represented by the
           dash-dotted (red) curve.}
  \label{fig:cos2t_8}
\end{figure}

In Fig.~\ref{fig:cos2t_8}, we demonstrate the temperature dependence of
the molecular response to the laser in the temperature range
from~$0 \U{K}$ to $20 \U{K}$.
The transition from the intermediate region of molecular alignment
to purely adiabatic alignment is observed:
at $T = 0 \U{K}$, the curve is much broader than the laser pulse and
non-Gaussian which exhibits the nonlinear response of~$\expectval{\cos^2 \vartheta}(t)$
with respect to the Gaussian laser pulse.
The rising temperature leads to a considerable reduction of the peak alignment.
Also the large oscillations beyond a time of~$150 \U{ps}$ vanish completely.
A broad thermal distribution is responsible for dephasing effects which
diminish nonadiabaticities.~\cite{Buth:LA-08}
For a given temperature, the degree of adiabaticity can be judged by applying
our criterion from Sec.~\ref{sec:thaverp}: we compare the FWHM laser-pulse
duration with the thermally averaged rotational period [see
Table~\ref{tab:thermave} for examples].
Generally, the values of~$T\I{th}$ in the table turn out to be smaller than~%
$T\I{RP} = 1 / (2B) = 238.8 \U{ps}$ for~\CFtBr{}, as they should, to reflect the
right trend: for small temperatures, the period is longer than the laser pulse, thus
correctly predicting a nonadiabatic molecular response in Fig.~\ref{fig:cos2t_8};
for larger temperatures an adiabatic behavior is rightly indicated.
The considerable sensitivity of the rotational response of molecules to the rotational
temperature suggests to use it to measure~$\expectval{\cos^2 \vartheta}(t)$, \ie,
the effect can be harnessed as a thermometer to determine the rotational
temperature of a gas sample.
However, one has to keep in mind that in experiments, the precise
determination of the laser intensity is often very difficult.

%
%
\begin{table*}
  \centering
  \begin{ruledtabular}
    \begin{tabular}{lccccccc}
      $T$ [K]        & 0        & 0.1 &  1 &  3 &  5 & 10 & 20 \\
      $T\I{th}$ [ps] & $\infty$ & 167 & 99 & 65 & 52 & 39 & 28 \\
    \end{tabular}
  \end{ruledtabular}
  \caption{Thermally averaged rotational periods~$T\I{th}$
           [Eq.~(\ref{eq:T_rot})] for a number of rotational temperatures~$T$
           for the linear rotor model of~\CFtBr{}.
           The periods were determined by using up to~$J=1000$ terms in the
           sum over~$J$ in Eqs.~(\ref{eq:T_rot}) and (\ref{eq:partition}).}
  \label{tab:thermave}
\end{table*}

Overlaid on the curves for the linear rotor are data from a symmetric
rotor computation.
The deviations between both descriptions of~\CFtBr{} are minute.
For $T=0$, there is no difference between both because only the
rotational ground state with~$J=0$ is populated.
As the interaction with the laser
does not mix states with different $K$~quantum numbers,~\cite{Buth:LA-08}
no difference may arise when a laser is turned on.
Also for higher~$T$, the deviation remain tiny.

\begin{figure}
  \includegraphics[clip,width=\hsize]{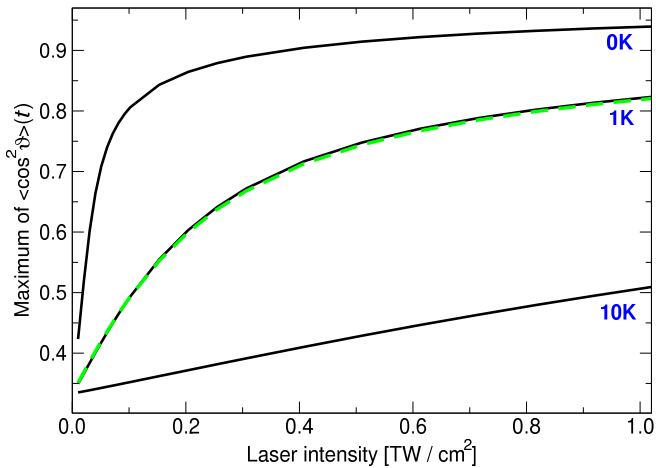}
  \caption{(Color online) Influence of the laser intensity on the maximally
           attainable alignment of~\CFtBr{}.
           We show curves for the linear rotor as solid (black) lines
           and curves for the symmetric rotor as green (dashed) lines.
           The data are for various temperatures given by the numbers in blue.
           We use laser pulses with a FWHM duration of~$\tau\I{L} = 1 \U{ns}$.}
  \label{fig:intensdep_4}
\end{figure}

In Fig.~\ref{fig:intensdep_4}, we investigate the influence of the laser intensity
on the maximally attainable alignment.
We take the maximum of~$\expectval{\cos^2 \vartheta}(t)$ over the propagation time
interval for a number of temperatures and laser intensities.
The computational parameters are chosen such that we are in the adiabatic regime.
For all the curves, we have the following two limits:
if the laser intensity is zero, there is no alignment, only a thermal
ensemble and $\expectval{\cos^2 \vartheta}(t) = \frac{1}{3}$.
Conversely, if $I\I{L,0} \to \infty$, then $\max \{\expectval
{\cos^2 \vartheta}(t)\} \to 1$.
Of course, the latter limit can never be approached closely in practice
because molecules are ionized above a certain value for the intensity.
With increasing temperature, the curves in Fig.~\ref{fig:intensdep_4} change
dramatically.
For $0 \U{K}$, we find a strong nonlinear dependence of~$\max \{\expectval
{\cos^2 \vartheta}(t)\}$ on the intensity.
For $20 \U{K}$, the dependence is essentially linear.
In analogy to the discussion of Fig.~\ref{fig:cos2t_8}, we conclude that
the strong temperature dependence of the functional shape of~$\max \{\expectval
{\cos^2 \vartheta}(t)\}$ provides a means to determine the rotational temperature
of a gas sample, if the laser intensity is accurately known.

\subsection{X-ray absorption}
\label{sec:xrabsi}

The expectation value~$\expectval{\cos^2 \vartheta}(t)$ examined in the
previous Sec.~\ref{sec:lasalign} can be inferred experimentally
via the Coulomb explosion technique.
However, due to the complicated nature of the interaction of a laser
with a molecule, this experimental technique is a somewhat intricate
way to study molecular alignment;
\eg, recently, the strong-field ionization probability of aligned
molecules was measured and a nontrivial, molecule-specific angular
dependence was found.~\cite{Pavicic:DM-07}
Therefore, this technique is prone to systematic errors~\cite{Ellert:MA-99}
and it is highly desirable to have an alternative
method for the measurement of molecular alignment.
We advocate resonant absorption of polarized x~rays as a novel route to
study molecular alignment.~\cite{Santra:SF-07,Buth:LA-08,Peterson:XR-08}

\begin{figure}
  \includegraphics[clip,width=\hsize]{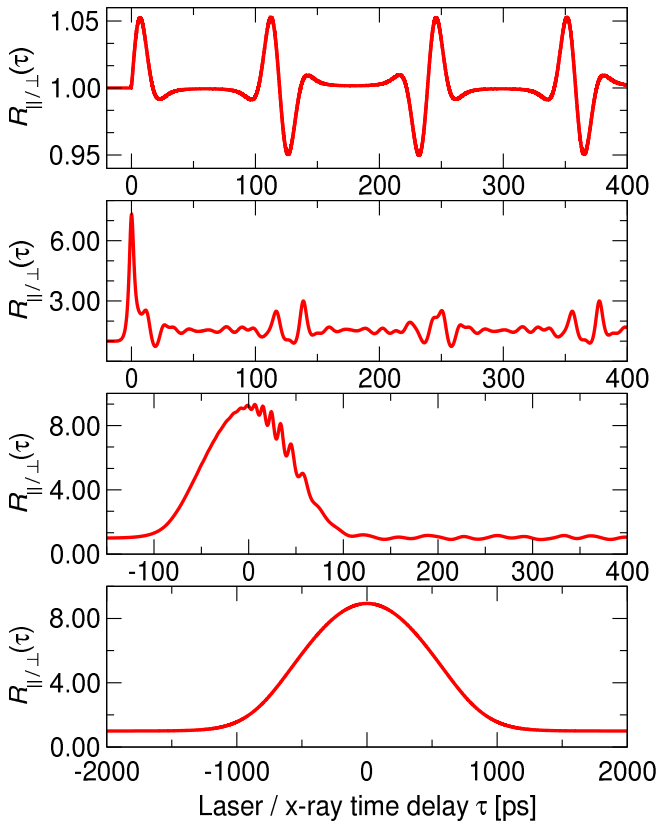}
  \caption{(Color online) Study of the four regimes of the alignment
           of~\CFtBr{} from Fig.~\ref{fig:cos2t_7} with a picosecond x-ray
           source ($\tau\I{X} = 1 \U{ps}$).
           The dependence of~$R\I{\parallel / \perp}(\tau)$
           [Eq.~(\ref{eq:crossparperp})] on the time delay~$\tau$ between
           laser and x-ray pulses is shown.}
  \label{fig:cc1ps}
\end{figure}

In Fig.~\ref{fig:cc1ps}, we study the four regimes of molecular alignment
from Fig.~\ref{fig:cos2t_7} using picosecond x-ray pulses
($\tau\I{X} = 1 \U{ps}$).
The x-ray pulse duration was chosen much shorter than the typical
rotational time scale~$T\I{RP}$.
We investigate the amount of information on the rotational dynamics which
is provided by the ratio~$R\I{\parallel / \perp}(\tau)$
[Eq.~(\ref{eq:crossparperp})] because it is an experimentally accessible
quantity which has been measured successfully before by
Peterson~\etal~\cite{Peterson:XR-08}
The dependence of~$R\I{\parallel / \perp}(\tau)$ [Eq.~(\ref{eq:crossparperp})]
on the laser pulse is shown for a large range of time delays~$\tau$ between laser
and x-ray pulses.
With the picosecond pulses, the molecular dynamics is reproduced clearly
in Fig.~\ref{fig:cc1ps};
even the detailed features of Fig.~\ref{fig:cos2t_7} are resolved.
This shows that short-pulse x-ray absorption is a promising novel avenue to
study the details of molecular rotations.

\begin{figure}
  \includegraphics[clip,width=\hsize]{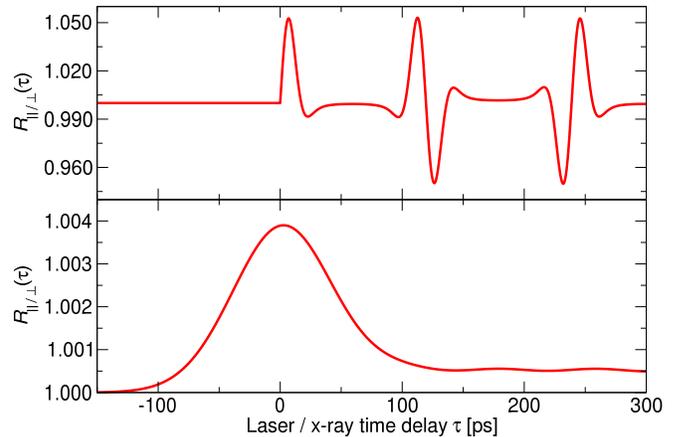}
  \caption{(Color online) Impulsive alignment of~\CFtBr{} studied by
           x-ray absorption.
           The setup is the one of the top panel in Figs.~\ref{fig:cos2t_7}
           and \ref{fig:cc1ps}.
           The ratio~$R\I{\parallel / \perp}(\tau)$
           [Eq.~(\ref{eq:crossparperp})] is shown here.
           We use~$\tau\I{X} = 50 \U{fs}$ (top) and
           $\tau\I{X} = 100 \U{ps}$ (bottom).
           Other computational parameters are~$T = 1 \U{K}$
           and $\tau\I{L} = 50 \U{fs}$.}
  \label{fig:cc50fs_100ps}
\end{figure}

In Fig.~\ref{fig:cc50fs_100ps}, we investigate the impact of the duration
of the x-ray pulse on the absorption $R\I{\parallel / \perp}(\tau)$
[Eq.~(\ref{eq:crossparperp})].
We find that the curve in the top panel of Fig.~\ref{fig:cc50fs_100ps}
for~$\tau\I{X} = 50 \U{fs}$ can hardly be distinguished from the curve
for~$R\I{\parallel / \perp}(\tau)$
in the top panel of Fig.~\ref{fig:cc1ps} for~$\tau\I{X} = 1 \U{ps}$.
This indicates that picosecond x-ray pulses are sufficiently short
to deliver snapshots without averaging noticeably over the
rotation of the molecule.
In other words, for~\CFtBr{} picosecond x-ray pulses can be modeled as a
$\delta(t - \tau)$~distribution~(\ref{eq:deltaflux}) and one
obtains~$r\I{\parallel / \perp}(\tau)$ [Eq.~(\ref{eq:crossparperpcross})].
In the lower panel of Fig.~\ref{fig:cc50fs_100ps}, we use long x-ray
pulses~$\tau\I{X} = 100 \U{ps}$, instead, to
determine~$R\I{\parallel / \perp}(\tau)$.
We observe that the detailed information on the molecular dynamics revealed
in the upper panel is basically invisible in the lower panel.
The long x-ray pulse averages out all fine details.

\begin{figure}
  \includegraphics[clip,width=\hsize]{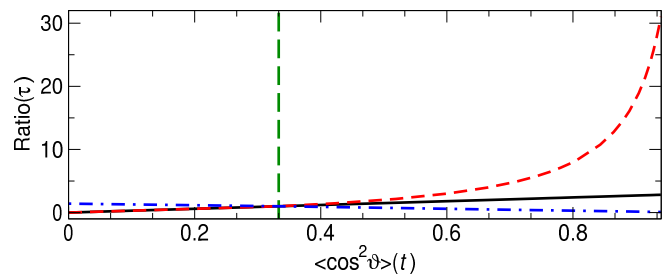}
  \caption{(Color online) The relation of~$\expectval{\cos^2
           \vartheta}(t)$ to the cross correlation
           ratios~$R\I{\parallel / \perp      }(\tau)$ (dashed red),
                  $R\I{\parallel / \mathrm{th}}(\tau)$ (solid black), and
                  $R\I{\perp     / \mathrm{th}}(\tau)$ (dash-dot blue)
           [Eq.~(\ref{eq:crossratios})] for~\CFtBr{}.
           The long dashed (green) vertical line indicates the value of a
           thermal ensemble.
           Other computational parameters are~$T = 0 \U{K}$,
           $\tau\I{L} = 1 \U{ns}$, and $\tau\I{X} = 10 \U{ps}$.}
  \label{fig:cos2t_ratio}
\end{figure}

Comparing the details of the curves in Fig.~\ref{fig:cc1ps} with the
corresponding curves from Fig.~\ref{fig:cos2t_7} reveals a vertical distortion.
Consider, for instance, the bottom panel in Fig.~\ref{fig:cos2t_7}.
While the curve is basically flat around the peak, the corresponding
curve in Fig.~\ref{fig:cc1ps} exhibits a clear peak.
The differences between the curves can be understood by examining the relation
between~$\expectval{\cos^2 \vartheta}(t)$ and $R\I{\parallel / \perp}(\tau)$
(we assume a $\delta$-distribution x-ray pulse~(\ref{eq:deltaflux}) and
equate~$t$ and $\tau$ here).
For this purposes, we carried out absorption calculations
for adiabatically aligned molecules for a series of laser intensities.
In Fig.~\ref{fig:cos2t_ratio}, we plot~$R\I{\parallel / \perp}(\tau)$
versus~$\expectval{\cos^2 \vartheta}(t)$.
To this end, we use the maximum of the two quantities over the propagation
interval of individual computations with varying laser intensity.
The curve reveals the nonlinear relationship between both quantities,
see Eq.~(\ref{eq:crossparperptwo}).
In transient alignment, \eg, the top panel of Fig.~\ref{fig:cos2t_7},
there is also a suppression of alignment with respect to the signal
of a thermal ensemble due to antialignment (\ie, alignment perpendicular
to the laser polarization axis).
We can easily add this part of the curve to Fig.~\ref{fig:cos2t_ratio} by realizing that
at~$\expectval{\cos^2 \vartheta}(t) = 0$, the molecules are perfectly antialigned.
According to Sec.~\ref{sec:elstructmod}, there is no absorption of x~rays
with a polarization parallel to the laser polarization axis.
We interpolate linearly between this point and the value unity for a
thermal ensemble.

In addition to a curve for~$R\I{\parallel / \perp}(\tau)$, we also show
results for~$R\I{\parallel / \mathrm{th}}(\tau)$ and $R\I{\perp / \mathrm{th}}(\tau)$
in Fig.~\ref{fig:cos2t_ratio}.
These two mappings of x-ray absorption ratios to~$\expectval{\cos^2 \vartheta}(t)$
are linear, see Eqs.~(\ref{eq:crossparthtwo}) and (\ref{eq:crossperpthtwo}).
They are opposite to each other: where the one is maximal, the other is minimal.
Yet the maximum of~$R\I{\perp / \mathrm{th}}(\tau)$ is only half the maximum
of~$R\I{\parallel / \mathrm{th}}(\tau)$.
This is due to the fact that there are two axes perpendicular to the laser
polarization axis but only one parallel to it.
The curves were extended into the range from~$\expectval{\cos^2
\vartheta}(t) = 0$ to the value of a thermal ensemble~$\expectval{\cos^2
\vartheta}(t) = \frac{1}{3}$ in the same way as for~$R\I{\parallel / \perp}(\tau)$.
For $R\I{\parallel / \mathrm{th}}(\tau)$, we have the two values~0 and 1, respectively.
The curve for~$R\I{\perp / \mathrm{th}}(\tau)$ certainly assumes the value unity for
a thermal ensemble as well.
However, in~$\expectval{\cos^2 \vartheta}(t) = 0$, we used the maximum value
of~$R\I{\parallel / \mathrm{th}}(\tau)$ divided by~2.
The signature of molecular alignment from the ratios~$R\I{\perp / \mathrm{th}}(\tau)$ and $R\I{\parallel / \mathrm{th}}(\tau)$
is weaker than from~$R\I{\parallel / \perp}(\tau)$.
The latter ratio is thus preferable in experiments.

\begin{figure}
  \includegraphics[clip,width=\hsize]{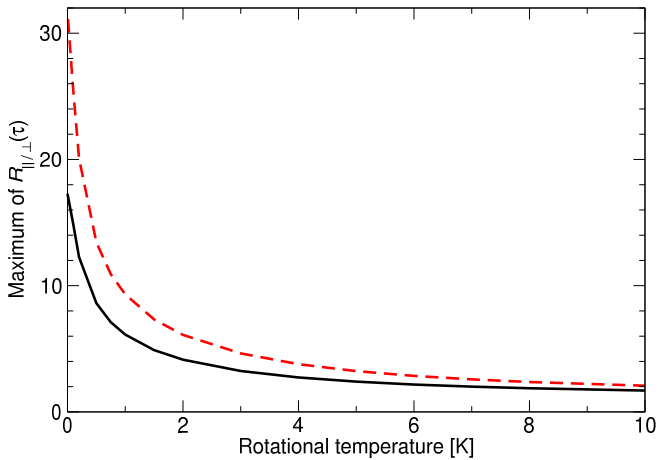}
  \caption{(Color online) Dependence of the x-ray absorption
           signal~$\max \{R\I{\parallel / \perp}(\tau)\}$
           [Eq.~(\ref{eq:crossparperp})]
           of~\CFtBr{} on the rotational temperature of the molecules.
           The solid (black) curve is for~$\tau\I{L} = 95 \U{ps}$ and
           $\tau\I{X} = 100 \U{ps}$ whereas the dashed (red) curve is
           for~$\tau\I{L} = 1 \U{ns}$ and $\tau\I{X} = 10 \U{ps}$.
           For both curves, the peak laser intensity is~$I\I{L,0}
           = 10^{12} \U{\frac{W}{cm^2}}$.}
  \label{fig:r1_160ps}
\end{figure}

In Fig.~\ref{fig:r1_160ps}, we investigate the dependence of the x-ray
absorption signal~$R\I{\parallel / \perp}(\tau)$ [Eq.~(\ref{eq:crossparperp})]
on the rotational temperature.
To this end, we computed the absorption of adiabatically aligned molecules
for a series of temperatures.
The high sensitivity of the molecular dynamics on the temperature revealed
in Fig.~\ref{fig:r1_160ps} parallels the one found in
Fig.~\ref{fig:cos2t_8}.
Our prediction stresses once more how important a low temperature is to
observe a substantial signal.
Conversely, the maximum of the x-ray absorption ratio can again be used to determine
the rotational temperature of a gas sample.
For this purpose, the precise laser and x-ray pulse parameters need to be
known to tailor a figure like Fig.~\ref{fig:r1_160ps} for a given experimental
situation.

\subsection{Control of x-ray absorption}

\begin{figure}
  \includegraphics[clip,width=\hsize]{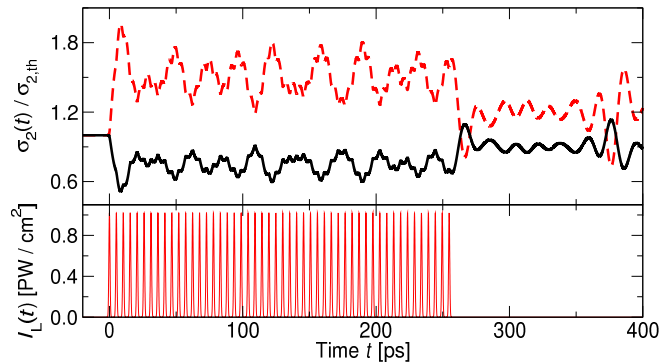}
  \caption{(Color online) Exemplary one-dimensional control of x-ray absorption
           by laser-aligned \CFtBr{}~molecules.
           In the upper panel, we show the induced molecular dynamics subject to
           the sequence of 50~laser pulses with intensity~$I\I{L}(t)$ which
           is shown in the lower panel.
           The individual pulses have a FWHM duration of~$\tau\I{L} = 1 \U{ps}$
           and are spaced by~$5.2 \U{ps}$.
           The curves in the upper panel represent instantaneous ratios of
           cross sections: $r\I{\parallel / \mathrm{th}}(t)$
           [Eq.~(\ref{eq:crossparthcross})]
           (dashed, red) and $r\I{\perp / \mathrm{th}}(t)$
           [Eq.~(\ref{eq:crossperpthcross})]
           (solid, black).
           The rotational temperature is~$1 \U{K}$.}
  \label{fig:control}
\end{figure}

In the previous Sec.~\ref{sec:xrabsi}, we used x-ray absorption to study the
rotational molecular dynamics of a gas sample.
In this subsection, we focus on the inverse process: the control of x-ray
absorption by molecular alignment.
Controlling x-ray pulse shapes represents a powerful tool to control
inner-shell electronic processes;
it has been theoretically established in terms of electromagnetically induced
transparency~(EIT) for x~rays.~\cite{Buth:TX-07,Buth:ET-07,Santra:SF-07,%
Buth:AR-08,Buth:RA-up,Buth:US-up}
There we use a shaped laser pulse to cut out two short x-ray pulses
from a long Gaussian x-ray pulse.
Here, we would like to investigate similar avenues using laser-aligned
molecules.
The control of molecular rotation with appropriately shaped laser pulses
has been investigated, \eg, in Ref.~\onlinecite{Bartels:PM-01,Pelzer:OC-07}.
The x-ray absorption cross section of the molecules depends on the alignment
[Sec.~\ref{sec:elstructmod}].
We assume that the gas sample absorbs most x~rays for randomly oriented
molecules.
Once the molecules are sufficiently antialigned, a large fraction of the
incident x-ray flux is transmitted through the gas sample for parallel
laser and x-ray polarization vectors.

Let us examine examples of controlled x-ray absorption.
In the upper panel Fig.~\ref{fig:control}, we show the ratios of cross
sections~$r\I{\parallel / \mathrm{th}}(t)$ [Eq.~(\ref{eq:crossparthcross})]
and $r\I{\perp / \mathrm{th}}(t)$ [Eq.~(\ref{eq:crossperpthcross})]
of~\CFtBr{} for parallel and
perpendicular x-ray and laser polarizations, respectively.
The molecules are subject to the sequence of 50~laser pulses shown in the
lower panel.
As derived in Sec.~\ref{sec:relcrosscos}, $\sigma_{2,\parallel}(t)$
[Eq.~(\ref{eq:xsecttwocospar})]
is proportional to~$\expectval{\cos^2 \vartheta}(t)$ and thus is well-suited
to examine the molecular motion.
The sequence imprints an oscillatory occurrence and vanishing of enhanced alignment
and antialignment.
This rotational motion is enforced by the laser and only present during the pulse
sequence.
Afterwards, $t > 260 \U{ps}$, we see---by comparison with the upper panel in
Fig.~\ref{fig:cos2t_7} (keep in mind that $\expectval{\cos^2 \vartheta}(t)$ is
plotted there)---that the free time-evolution after a single pulse is
approximately assumed.
The ratios~$r\I{\parallel / \mathrm{th}}(t)$
and $r\I{\perp / \mathrm{th}}(t)$
are complementary:
the former peaks where the latter is minimal.

Likewise, the ratios of cross sections corresponding to the physical situation
in the top panel of Fig.~\ref{fig:cos2t_7} can be investigated where
a single, short laser pulse creates a rotational wavepacket.
About every $120 \U{ps}$, the molecule undergoes short time intervals of
transient antialignment.
The brief periods of antialignment then lead to brief periods of transparency
of the sample for x~rays, \ie, x-ray bursts of a few picoseconds may be cut
out of a longer x-ray pulse.
We have constructed a picosecond clock which provides us with short
x-ray bursts.
For very long x-ray pulses, several bursts with a well-defined recurrence time
can be generated.

Control of x-ray absorption with laser-aligned molecules has advantages and
disadvantages compared with the competing technique of
EIT for x~rays.~\cite{Buth:TX-07,Buth:ET-07,Santra:SF-07,Buth:AR-08,%
Buth:RA-up,Buth:US-up}
An advantage of molecular alignment over EIT is that by the choice of the
molecule, x-ray absorption can be controlled for a large number of
wavelengths, \ie, by replacing~\CFtBr{} with other molecules,
one can shift the absorption edge over a wide range of energies.
Additionally, other edges, $L$, $M$, $N$, $O$, $P$, $Q$, can be used
alternatively without any extra effort.
This adds an additional degree of freedom of available x-ray wavelengths.
In the case of EIT for x~rays, on the other hand, the number of suitable
atoms seems to be quite limited by the requirements to sustain strong
laser fields and the availability of intense laser systems with the right
wavelength to couple two empty Rydberg orbitals.
Also with molecular alignment, absorption can be controlled for
hard x~rays where $K$-shell core-hole decay widths are large and EIT for
x~rays is suppressed.~\cite{Buth:TX-07,Buth:ET-07,Santra:SF-07,%
Buth:AR-08,Buth:RA-up,Buth:US-up}

%
%
%
%
Let us estimate the required length of \CFtBr{}~gas where the laser and the x~rays
need to overlap for the proposed x-ray pulse shaping to work.
We approximate the cross section of \CFtBr{} on the pre-edge resonance
by the $1s \to 5p$~cross section for krypton atoms~$\sim 17 \U{kb}$.~\cite{Buth:TX-07}
Let the gas jet have a number density of~$5 \E{14} \U{cm^{-3}}$ as it had in the
experiment of Ref.~\onlinecite{Peterson:XR-08}.
From Beer's law,~\cite{Als-Nielsen:EM-01} we estimate that the overlap region
needs to be~$1 \U{km}$~long for the x-ray flux to drop by a factor
of~$1 / \euler$.
Clearly, this is an absurd requirement in practice.
A number density of~$7 \E{19} \U{cm^{-3}}$ would be required to reduce the length
of the overlap region to~$0.8 \U{cm}$.
Yet the number density in the gas jet cannot be increased indefinitely because the molecules
start to interact strongly with each other forming clusters and, eventually,
a condensed phase.
The x-ray cross section assumed in this estimate holds for perfectly aligned
molecules parallel to the x-ray polarization direction.
With laser-alignment, this cross section can be reduced to zero for perfectly
antialigned molecules.
Note that the cross section is about a hundred times
larger for neon atoms than for krypton atoms.~\cite{Buth:ET-07}
The estimate with respect to the fluorine pre-edge resonance is thus far more favorable
than for bromine.
On the down side of molecular alignment over EIT is that the time scales
of the former is about three orders of magnitude slower (picoseconds)
than the latter (femtoseconds).
The molecules need to have a low rotational temperatures, too;
it is not yet clear whether this is physically possible because interactions between
molecules have a considerable damping influence on molecular rotations in denser
gases and condensed matter.~\cite{Pelzer:OC-07}
The latter points seem to represent a big, if not insurmountable, obstacle
for control of x-ray absorption by laser-manipulated molecular rotation.
However, the ideas presented in this subsection might nevertheless
prove fruitful in conjunction with recent work on the laser-alignment of
molecules in solution.~\cite{Ramakrishna:IL-05}

\section{Conclusion}
\label{sec:conclusion}

This study is devoted to laser-induced rotational dynamics of
bromotrifluoromethane~(\CFtBr{})~molecules probed by x~rays.
First, we investigated the molecular rotations in terms of the expectation
value~$\expectval{\cos^2 \vartheta}(t)$.
Second, we investigate the properties of a probe of molecular rotations
by x~rays.
This represents an alternative to the established Coulomb-explosion
technique.~\cite{Stapelfeldt:AM-03}
Third, we explore possibilities to employ laser manipulation of molecular
rotation to control x-ray pulse shapes, \ie, to imprint a certain pulse shape onto
a longer x-ray pulse.

We start by investigating~$\expectval{\cos^2 \vartheta}(t)$.
Specifically, we examine its temporal evolution for varying laser-pulse durations,
its temperature dependence, and its dependence on the peak laser intensity.
Then, we revisit these physical situations with short-pulse x~rays.
Additionally, we explore the dependence on the x-ray pulse duration and the
relation between the maximum of x-ray absorption and the maximum
of~$\expectval{\cos^2 \vartheta}(t)$.
We show that~$\expectval{\cos^2 \vartheta}(t)$ can be directly measured
by short-pulse x-ray absorption measurements.
Such short pulses will be produced by the emerging ultrafast x-ray
sources.~\cite{LCLS:CDR-02,%
Tanaka:SC-05,Borland:SA-05,Altarelli:TDR-06}
The method permits to stroboscopically photograph the steps in the time-evolution
of rotational wavepackets.
We can learn about the beams and the experimental conditions;
x-ray absorption is an \emph{in situ} probe.

Our investigations are based on our recent theory~\cite{Buth:LA-08} for the
laser alignment of symmetric-top molecules probed by x~rays.
To carry out the computations in this paper, we devise a parameter-free
two-level model for the electronic structure of~\CFtBr{};
it models the Br$\,1s \to \sigma^*$~pre-edge resonance.
The x-ray absorption cross section formula from Ref.~\onlinecite{Buth:LA-08}
is reduced to this model.
We employ coupled-cluster linear-response techniques to determine the
average dynamic dipole polarizability and the dynamic dipole polarizability
anisotropy for~\CFtBr{} in the laser light.
Finally, the usual temperature-independent criterion of adiabaticity for the laser
pulse duration~$\tau\I{L} \gg \frac{1}{2B}$ looses its predictive power because
the rotational response becomes more and more adiabatic with increasing rotational
temperature.
Therefore, we devise a thermally averaged rotational period as a new adiabaticity
criterion for rotational dynamics.

Based on our work, a number of future prospects offer themselves.
Absolute cross sections can be easily determined parametrizing the two-level
model fully with the cross section on the Br$\,1s \to \sigma^*$~pre-edge resonance
for molecules perfectly aligned along the x-ray polarization axis.
This parameter can be experimentally determined for clamped molecules
on surfaces~\cite{Stohr:NE-96} or computationally using \emph{ab initio}
methods.

Our rigid-rotor model has a few limitations.
No vibrations or internal rotations were accounted for.
In future work, such effects beyond the rigid-rotor approximation should be
addressed.
Additionally, our formalism can be extended to three-dimensional alignment
of asymmetric-top molecules.~\cite{Stapelfeldt:AM-03,Seideman:NA-05}
This enables one to study a large class of molecules which basically
do not rotate.

Our proposal to use laser-aligned molecules to control x-ray pulse shapes
offers a higher flexibility in terms of the suitable x-ray energies than
electromagnetically induced transparency for
x~rays which has been studied only for rare-gas atoms so far.~\cite{Buth:TX-07,%
Buth:ET-07,Santra:SF-07,Buth:AR-08,Buth:RA-up,Buth:US-up}
Chemical shifts of the inner-shell edges offer an added flexibility to
manipulate the resonance energies of molecules compared with atoms.
To control x~rays by laser-aligned molecules in practice, one has to
experimentally produce a cold beam of molecules with sufficiently high
number density.
However, at present, this requirement seems to inhibit a success of this
technique both in principle and in practice.
Further, one has to make theoretical advancements to predict the signal of
such experiments.
At the required number densities, one faces condensed-matter;
the interactions with neighboring molecules are strong
and need to be incorporated in terms of a more comprehensive
theory.~\cite{Ramakrishna:IL-05}
The laser pulses need to be shaped such that a desired rotational dynamics
is imposed on the molecules.
Approaches to the quantum control of molecular alignment have been derived,
\eg, Refs.~\onlinecite{Bartels:PM-01,Pelzer:OC-07} offering
perspectives for an optimal control of the rotational dynamics of molecules
and, in this way, an optimal, ultrafast (picoseconds) control of x-ray pulse
shapes.

\begin{acknowledgments}
We would like to thank Linda Young for fruitful discussions.
C.B.'s research was partly funded by a Feodor Lynen Research Fellowship from
the Alexander von Humboldt Foundation.
C.B.'s and R.S.'s work was supported by the Office of Basic Energy Sciences,
Office of Science, U.S.~Department of Energy, under Contract
No.~DE-AC02-06CH11357.
\end{acknowledgments}

\appendix*
\section{Symmetry-breaking effects in \protect\CFtBr}

In Sec.~\ref{sec:elstructmod} we argued that because of the C$_{3v}$
symmetry of \CFtBr{}, the transition dipole vector between the
Br$\,1s$ and the $\sigma^*$~orbitals has only a nonzero $c$-component.
Two basic mechanisms can lead to a failure of this prediction.

The first mechanism is symmetry breaking induced by molecular vibrations.
We consider this effect to be negligible for the following reason.
Laser-induced alignment in the gas phase requires rotationally cold molecules.
We may therefore assume that before absorbing an x-ray photon, the molecules
are in the electronic and vibrational ground state.~\cite{Seideman:NA-05}
Hence, it is only vibrations in the core-excited resonance state that
could break the C$_{3v}$ symmetry of \CFtBr{}.
Since a $1s$~vacancy in a bromine atom has a decay width
of~\cite{Campbell:WA-01}~$2.5 \U{eV}$ and the chemical environment has
only a negligible influence on this core-hole decay, the lifetime of a
Br$\,1s \to \sigma^*$~resonance state of~\CFtBr{} can be inferred to be
only~$\frac{1}{\Gamma} = 0.26 \U{fs}$.
This time is insufficient for symmetry-breaking deformations.

The second mechanism is symmetry breaking induced by spin-orbit coupling
in bromine.
To estimate the impact of this effect, we employ a simple model based on
the following four classes of normalized atomic basis
functions~$\ket{\chi_i,m_s} = \ket{\chi_i} \otimes \ket{m_s}$
with~$i = 1, 2, 3, 4$.
We denote by~$\ket{m_s}$ a Pauli spinor with spin projection quantum
number~$m_s=\pm 1/2$.
Here, $\ket{\chi_1}$~represents a $sp^3$~hybrid orbital centered on carbon
and directed towards bromine.
The Br$\,4p$~orbitals are~$\ket{\chi_2}$, $\ket{\chi_3}$, $\ket{\chi_4}$
with orbital angular momentum projection quantum numbers~$m_l=0, 1, -1$,
respectively.

Let $\hat{F}$ stand for the spin-orbit-free Fock operator associated with
the closed-shell electronic ground state of~\CFtBr{} at equilibrium geometry.
Molecular orbitals~$\ket{\varphi_i,m_s}$---in the absence of spin-orbit
coupling---have the spin-independent orbital energies~$\varepsilon_i$.
We assume that the molecular orbitals of interest may be
written as superpositions of the atomic orbitals $\ket{\chi_i,m_s}$.
Specifically, for the doubly occupied $\sigma$~orbital (HOMO-1), $i = 1$,
and the unoccupied $\sigma^{\ast}$~orbital (LUMO), $i = 2$,
we write
\begin{equation}
  \label{A6}
  \ket{\varphi_i,m_s} = c_{1,i} \ket{\chi_1,m_s} + c_{2,i} \ket{\chi_2,m_s} \; ;
\end{equation}
the degenerate, doubly occupied lone pairs (HOMO)~$i = 3, 4$ are
approximated by
\begin{equation}
  \label{A8}
  \ket{\varphi_i,m_s} = \ket{\chi_i,m_s} \; .
\end{equation}

We form the effective one-electron Hamiltonian~$\hat H\I{eff} = \hat F
+ \hat V\I{SO}$ with the spin-orbit coupling~$\hat
V\I{SO}$~\cite{Rose:ET-57,Strange:RQ-98,Merzbacher:QM-98}.
Using first-order perturbation theory, the $\sigma^{\ast}$~orbital
in the presence of spin-orbit coupling is
\begin{equation}
  \label{A11}
  \begin{array}{rl}
    \displaystyle \ket{\tilde{\varphi}_2,m_s} =& \displaystyle
      \ket{\varphi_2,m_s} \\
  &\displaystyle {}+ \sum^4_{\scriptstyle i=1 \atop \scriptstyle i\ne 2}
    \sum_{m'_s = -\frac{1}{2}}^{\frac{1}{2}}
    \frac{\bra{\varphi_i,m'_s} \hat V\I{SO} \ket{\varphi_2,m_s}}
         {\varepsilon_2 - \varepsilon_i} \\
  &\displaystyle \hspace{10em} {}\times \ket{\varphi_i,m'_s} \; .
  \end{array}
\end{equation}
The notation~$\ket{\tilde{\varphi}_2,m_s}$ shall not imply that~$m_s$
is a good quantum number;
it merely indicates that $\ket{\tilde{\varphi}_2,m_s}$
becomes~$\ket{\varphi_2,m_s}$ in the limit of vanishing spin-orbit
interaction.

Spin-orbit coupling is associated with the atomic potential near the bromine
nucleus.
We therefore assume that the $sp^3$ hybrid orbital on carbon is unaffected by
spin-orbit coupling and set
\begin{equation}
  \label{A12}
  \bra{\chi_1,m_s} \hat V\I{SO} \ket{\chi_i,m'_s} = 0 \; ,
\end{equation}
with~$i = 1, 2, 3, 4$ and $m_s, \  m'_s = -\frac{1}{2}, \frac{1}{2}$.
However, $\hat V\I{SO}$ causes mixing between the Br$\,4p$
orbitals~$\ket{\chi_2,m_s}$, $\ket{\chi_3,m_s}$, and $\ket{\chi_4,m_s}$.
For an atom, $\hat V\I{SO}$ is proportional
to~\cite{Rose:ET-57,Strange:RQ-98,Merzbacher:QM-98}~$\hat{\vec{l}} \cdot
\hat{\vec{s}}$, where $\hat{\vec{l}}$ and $\hat{\vec{s}}$ are the orbital
and spin angular momentum operators.
Thus, $\hat V\I{SO}$ is diagonal with respect to the spin-orbit coupled
states~\cite{Rose:ET-57,Strange:RQ-98}
\begin{equation}
  \label{A13}
  \begin{array}{rl}
    \ket{4p_j,m} =& \Sum_{m_l=-1\vphantom{\frac{1}{2}}}^{1
      \vphantom{\frac{1}{2}}} \Sum_{m_s=-\frac{1}{2}}^{\frac{1}{2}}
      \cleb{1,1/2,j}{m_l,m_s,m} \\
    &\hspace{6em} {}\times \ket{4p_{m_l}} \otimes \ket{m_s} \; .
  \end{array}
\end{equation}
In this expression, $\cleb{1,1/2,j}{m_l,m_s,m}$ is a Clebsch-Gordan
coefficient,~\cite{Rose:ET-57} $j=1/2$ or $3/2$, and $m=-j,\ldots,j$.
Upon inverting Eq.~(\ref{A13}),~\cite{Rose:ET-57} we obtain for the spin-orbit
coupling matrix elements in the atomic basis
\begin{equation}
  \label{A14}
  \begin{array}{rl}
    & \displaystyle \bra{4p_{m_l}} \otimes \bra{m_s} \hat V\I{SO}
      \ket{4p_{m'_l}} \otimes \ket{m'_s} \\
    \displaystyle =& \displaystyle \delta_{m_l+m_s,m'_l+m'_s}
      \Sum_{j=\frac{1}{2}}^{\frac{3}{2}} \cleb{1,1/2,j}{m_l,m_s,m_l+m_s} \\
    & \displaystyle {} \times \cleb{1,1/2,j}{m'_l,m'_s,m_l+m_s}
      \> \Delta E_{4p,j},
  \end{array}
\end{equation}
where~\cite{Merzbacher:QM-98}
\begin{equation}
  \label{A15}
  \Delta E_{4p,j} = \frac{\Delta E^{\mathrm{SO}}_{4p}}{3} \times
  \cases{
    \phantom{-}1 & , $j=3/2$ \cr
              -2 & , $j=1/2$ }
  \; ,
\end{equation}
and $\Delta E^{\mathrm{SO}}_{4p}$ is the fine-structure splitting in the
valence shell of atomic bromine.
Expression~(\ref{A14})---in combination with Eqs.~(\ref{A6}), (\ref{A8}), and
(\ref{A12})---justifies that terms involving matrix elements of the form
$\bra{\varphi_2,m_s} \hat V\I{SO} \ket{\varphi_2,m'_s}$ for~$m_s \ne m'_s$
were excluded from Eq.~(\ref{A11}).

Hence, upon collecting results, the spin-orbit coupled $\sigma^{\ast}$~orbital
may be written in a transparent form:
\begin{equation}
  \begin{array}{rcl}
    \slabel{A16}
    \displaystyle \ket{\tilde{\varphi}_2,\pm 1/2} &=& \displaystyle
      \ket{\varphi_2,\pm 1/2} \nonumber \\
    &&{}\displaystyle + c_{2,2} \frac{\sqrt{2}}{3} \frac{\Delta
      E^{\mathrm{SO}}_{4p}}{E\I{gap}} \ket{\varphi_3,\mp 1/2} \; ,
      \qquad\qquad
  \end{array}
\end{equation}
We have introduced~$E\I{gap} = \varepsilon_3 - \varepsilon_2$ using
Eqs.~(\ref{A6}) and (\ref{A8}) to denote the HOMO-LUMO gap.
Employing standard angular momentum algebra~\cite{Zare:AM-88} and exploiting
the fact that the Br$\,1s$ orbital $\ket{1s,m_s}$ has practically no overlap with
the $sp^3$~hybrid orbital on carbon, the x-ray absorption probability at the
Br$\,1s \rightarrow \sigma^{\ast}$ resonance is proportional to
\begin{equation}
  \label{A18}
  \begin{array}{rl}
    &\displaystyle \Sum_{m_s, m'_s = -\frac{1}{2}}^{\frac{1}{2}} \left|
      \bra{\tilde{\varphi}_2,m_s} \hat{\vec d} \cdot \vec e\I{X}
      \ket{1s,m'_s} \right|^2 \\
    = & \displaystyle \frac{2}{3} \> |c_{2,2}|^2 \; |\bra{4p}\!|\hat d
      |\!\ket{1s}|^2 \\
    & \displaystyle \times \biggl[ e\I{X,c}^2 +
      \frac{(\Delta E^{\mathrm{SO}}_{4p})^2}{9 \, E^2\I{gap}} \>
      [e\I{X,a}^2 + e\I{X,b}^2] \biggr] \; .
  \end{array}
\end{equation}
Here, $\hat{\vec d}$~is the electric dipole operator, $\vec e\I{X}$~is the x-ray
polarization vector, $\bra{4p}\!|\hat d|\!\ket{1s}$ is a reduced matrix
element,~\cite{Zare:AM-88} and $e\I{X,a}$, $e\I{X,b}$, and
$e\I{X,c}$ are the Cartesian components
of~$\vec e\I{X}$ in the body-fixed frame, assuming that the molecular
symmetry axis is the $c$~axis.

X-ray polarization parallel to the \CFtBr{} symmetry axis thus corresponds
to~$e\I{X,a}^2 + e\I{X,b}^2 = 0$ and $e\I{X,c}^2 = 1$;
x-ray polarization perpendicular to the \CFtBr{} symmetry axis corresponds
to~$e\I{X,a}^2 + e\I{X,b}^2 = 1$ and $e\I{X,c}^2 = 0$.
It follows from Eq.~(\ref{A18}) that, assuming perfect alignment of the
molecule relative to the x-ray polarization axis, the ratio between x-ray
absorption in the parallel configuration~($\parallel$) and x-ray absorption
in the perpendicular configuration~($\perp$) is given by
\begin{equation}
  \label{A19}
  f_{\parallel/\perp} = \frac{9 \, E^2\I{gap}} {(\Delta
    E^{\mathrm{SO}}_{4p})^2} \; .
\end{equation}
The spin-orbit splitting~$\Delta E^{\mathrm{SO}}_{4p}$ in a bromine atom
is~$0.46 \eV$.~\cite{NIST-3.1.5}
Using \textsc{dalton}~\cite{dalton:pgm-05} and the aug-cc-pVTZ basis
set,~\cite{Wilson:GBS-99,basislib:02-02-06} we find that the HOMO-LUMO
gap~$E\I{gap}$ in~\CFtBr{} is $14 \eV$.
We may therefore conclude from Eq.~(\ref{A19}) that for perfect alignment,
$f_{\parallel/\perp} \approx 8000$.
This upper limit is much higher than any of the values
for~$f_{\parallel/\perp}$ calculated in Sec.~\ref{sec:results}.
In comparison to the restrictions imposed on~$f_{\parallel/\perp}$ by
laser-induced alignment, the impact of spin-orbit coupling is negligible.


\begin{thebibliography}{52}
\bibitem[{foo({\natexlab{a}})}]{footnote2}
\bibinfo{note}{The molecular structure (Fig.~\ref{fig:geomCF3Br}) and the
  orbital plots (Fig.~\ref{fig:twoorbs}) of~CF$_3$Br were produced by the
  \textsc{molden}~\cite{Schaftenaar:MO-00} program based on a Hartree-Fock
  computation with \textsc{dalton}~\cite{dalton:pgm-05} using the
  cc-pVDZ~\cite{Wilson:GBS-99,basislib:02-02-06} basis set.}
\bibitem[{foo({\natexlab{b}})}]{footnote1}
\bibinfo{note}{The complex conjugate of the dipole vector is taken to suppress
  the complex conjugation in the Hermitian scalar product. In other words, the
  first scalar product in Eq.~(\ref{eq:xsect}) means~$\vec
  d^{\>\prime\,*}_{0i''} \mul \vec s_{J'KM, J''K''M''}^{\>\prime} =
  \Sum_{j=1}^3 (\vec d_{0i''}^{\>\prime})_j \, (\vec s_{J'KM,
  J''K''M''}^{\>\prime})_j$. Please note further that the definition~$\vec
  d^{\> \prime}_{0i''} = \bra{0} \hat{\vec d}\vphantom{\vec d}^{\> \prime\,*}
  \ket{i''}$ involves, in the general case, the complex conjugated dipole
  operator in the spherical basis. In Eq.~(33) of Ref.~\onlinecite{Buth:LA-08},
  a real dipole vector was assumed, \ie, only a vector with a $m =
  0$~component, which is the case for the molecules~Br$_2$ and \CFtBr{}.}
\bibitem[{foo({\natexlab{c}})}]{footnote3}
\bibinfo{note}{There are two stable, naturally occurring isotopes of bromine:
  $^{79}$Br and $^{81}$Br. We use only the most abundant isotope, $^{79}$Br, to
  determine the rotational
  constants.~\cite{Mills:QU-88,Rosman:IC-98,Buth:LA-08}}.
\bibitem[{foo({\natexlab{d}})}]{footnote4}
\bibinfo{note}{In mathematical terms, this means that only~$B$ enters the
  equation of motion of the rotational wavepacket driven by the laser [see
  Eq.~(26) in Ref.~\onlinecite{Buth:LA-08}].}
\bibitem[{foo({\natexlab{e}})}]{footnote5}
\bibinfo{note}{We define the end of the laser to be one FWHM
  duration~$\tau\I{L}$ (and farther) away from its peak.}
\end{thebibliography}
\end{document}